\pdfoutput=1

\documentclass[11pt,twoside,a4paper,cmspaper,final,collab]{cms-tdr}
\def\svnVersion{731f971-D}\def\svnDate{2019/12/09}\def\cmsCernNoTag{CERN-EP-2019-185}\def\cmsCernDate{\today}\def\cmsMessage{"Published in Physical Review D as \href{http://dx.doi.org/10.1103/PhysRevD.100.112003}{\doi{10.1103/PhysRevD.100.112003}.}"}

\begin{document}\cmsNoteHeader{EXO-19-005}

\hyphenation{had-ron-i-za-tion}
\hyphenation{cal-or-i-me-ter}
\hyphenation{de-vices}
\newcommand{\ctau}{\ensuremath{c\tau}\xspace}
\newcommand{\smaj}{\ensuremath{S_{\text{major}}}\xspace}
\newcommand{\smin}{\ensuremath{S_{\text{minor}}}\xspace}
\newcommand{\spp}{\ensuremath{S_{\phi\phi}}\xspace}
\newcommand{\see}{\ensuremath{S_{\eta\eta}}\xspace}
\newcommand{\sep}{\ensuremath{S_{\eta\phi}}\xspace}
\newcommand{\tECAL}{\ensuremath{t_{\text{ECAL}}}\xspace}
\newcommand{\tECALi}{\ensuremath{\tECAL^{i}}\xspace}
\newcommand{\Ai}{\ensuremath{A_{i}}\xspace}
\newcommand{\Aeff}{\ensuremath{A_{\text{eff}}}\xspace}
\newcommand{\sigN}{\ensuremath{\sigma_{\text{N}}}\xspace}
\newcommand{\AeosN}{\ensuremath{\Aeff/\sigN}\xspace}
\newcommand{\AsigNX}[1]{\ensuremath{(A_{#1}/\sigma_{\text{N}_{#1}})}\xspace}
\newcommand{\sigNi}{\ensuremath{\sigma_{\text{N}_{i}}}\xspace}
\newcommand{\sigSQi}{\ensuremath{\sigma^{2}_{i}}\xspace}
\newcommand{\tpho}{\ensuremath{t_{\PGg}}\xspace}
\newcommand{\onepho}[1]{\ensuremath{#1\PGg}\xspace}
\newcommand{\twopho}[1]{\ensuremath{#1\PGg\PGg}\xspace}
\newcommand{\gjets}{\PGg{}\text{+jets}\xspace}
\newcommand{\wjets}{\PW{}\text{+jets}\xspace}
\newcommand{\zjets}{\PZ{}\text{+jets}\xspace}
\newcommand{\NX}[1]{\ensuremath{N_{\text{#1}}}\xspace}
\newcommand{\Nobs}{\ensuremath{N_{\text{obs}}^{\text{data}}}\xspace}
\newcommand{\Npostfit}{\ensuremath{N_{\text{bkg}}^{\text{postfit}}}\xspace}
\newcommand{\Npostfitmask}{\ensuremath{N_{\text{bkg(no C)}}^{\text{postfit}}}\xspace}
\newcommand{\rXA}[1]{\ensuremath{r_{\text{#1/A}}}\xspace}
\newcommand{\Zepem}{\ensuremath{\PZ\to\Pep\Pem}\xspace}
\newcommand{\m}{\unit{m}}
\newcommand{\ns}{\unit{ns}}
\newcommand{\ps}{\unit{ps}}
\newcommand{\kHz}{\unit{kHz}}
\newcolumntype{C}[1]{>{\centering\arraybackslash}m{#1}}
\newcolumntype{L}[1]{>{\raggedright\arraybackslash}m{#1}}
\newlength\cmsTabSkip\setlength{\cmsTabSkip}{2ex}

\cmsNoteHeader{EXO-19-005}
\title{Search for long-lived particles using delayed photons in proton-proton collisions at \texorpdfstring{$\sqrt{s}=13\TeV$}{sqrt(s) = 13 TeV}}

\date{\today}

\abstract{
A search for long-lived particles decaying to photons and weakly interacting particles, using
proton-proton collision data at $\sqrt{s}=13\TeV$ collected by the CMS experiment
in 2016--2017 is presented. The data set corresponds to an integrated luminosity of 77.4\fbinv.
Results are interpreted in the context of supersymmetry with gauge-mediated supersymmetry
breaking, where the neutralino is long-lived and
decays to a photon and a gravitino. Limits are presented as a function
of the neutralino proper decay length and mass.
For neutralino proper decay lengths of 0.1, 1, 10, and
100\m, masses up to 320, 525, 360, and 215\GeV
are excluded at 95\% confidence level, respectively.
We extend the previous best limits in the neutralino proper decay length
by up to one order of magnitude, and in the neutralino mass by up to 100\GeV.
}

\hypersetup
{
pdfauthor={CMS Collaboration},
pdftitle={Search for long-lived particles using delayed photons in proton-proton collisions at sqrt(s)=13 TeV},
pdfsubject={CMS},
pdfkeywords={CMS, physics, photons, ECAL, time, ptmiss, GMSB}
}

\maketitle

\section{Introduction}
\label{sec:intro}

\begin{figure*}[hbtp]
\centering
\includegraphics[width=0.32\textwidth]{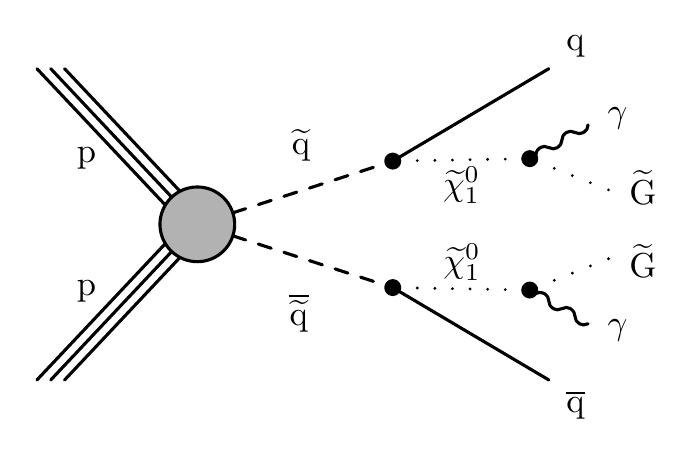}
\includegraphics[width=0.32\textwidth]{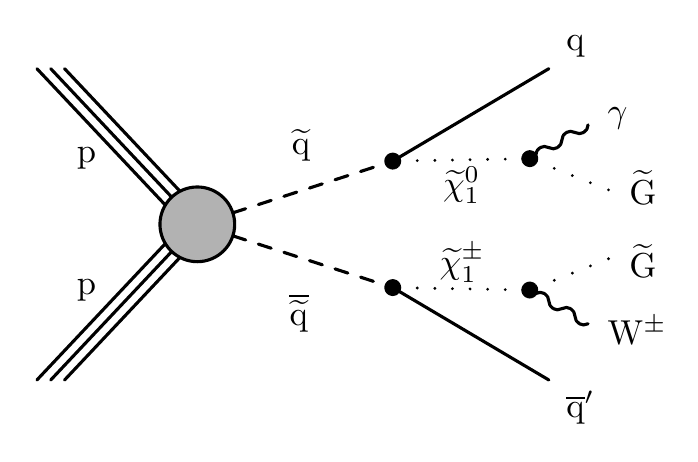}
\includegraphics[width=0.32\textwidth]{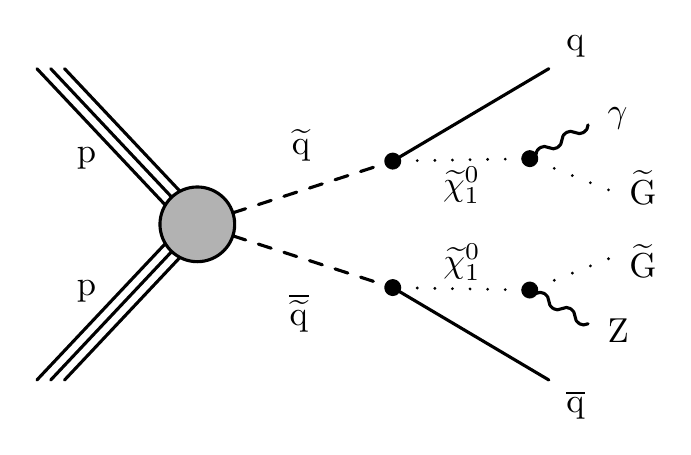}\\
\includegraphics[width=0.32\textwidth]{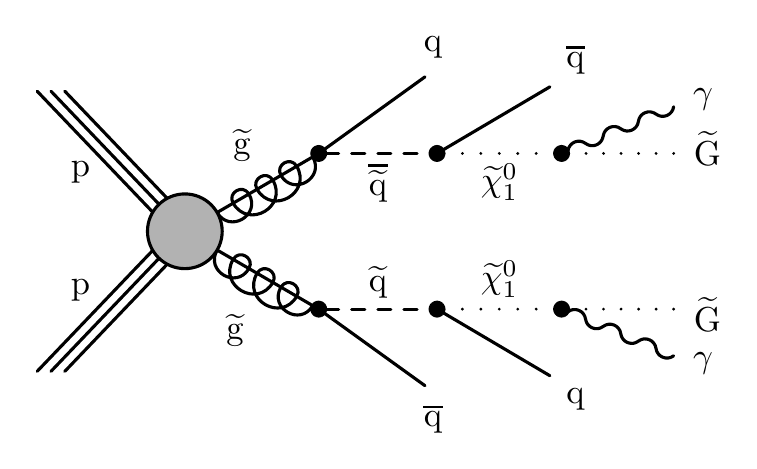}
\includegraphics[width=0.32\textwidth]{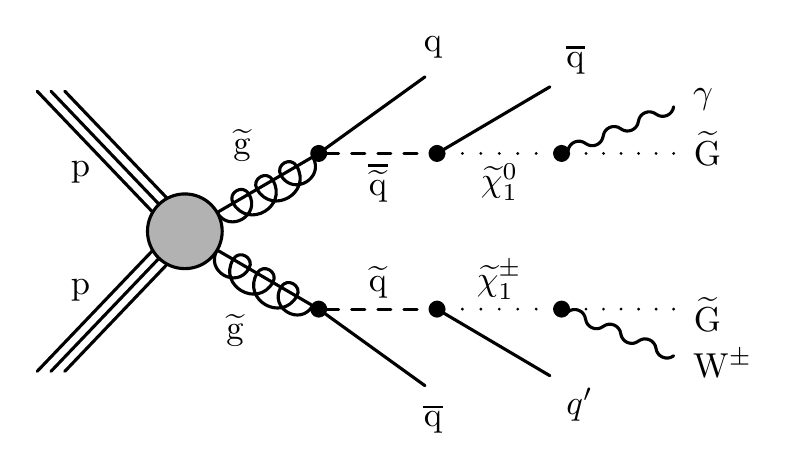}
\includegraphics[width=0.32\textwidth]{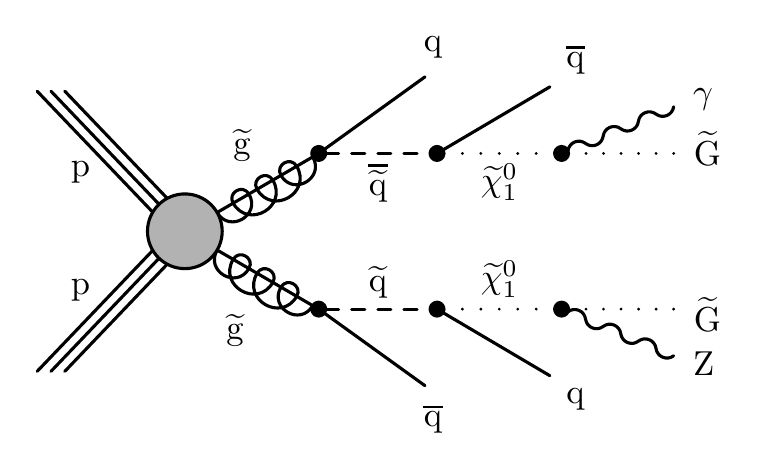}
\caption{Example Feynman diagrams for SUSY processes that result in diphoton (left) and single photon (middle and right) final states via squark (upper) and gluino (lower) pair-production at the LHC.}
\label{fig:feyn}
\end{figure*}

The results of a search for long-lived particles (LLP) decaying
to a photon and a weakly-interacting particle are presented.
Neutral particles with long lifetimes are predicted in many models of physics beyond
the standard model (SM). In this paper, a benchmark scenario of
supersymmetry (SUSY)~\cite{Ramond,Ramond:1971kx,Golfand,Volkov,Wess:1974tw,Freedman:1976xh,Deser:1976eh,Freedman:1976py,Ferrara:1976fu,Fayet,Chamseddine,Barbieri,Hall,Kane}
with gauge-mediated SUSY breaking (GMSB)~\cite{Giudice:1998bp,Dimopoulos:1996vz,GGMa,GGMd2,GGMd3,GGMd4,GGMd5,GGMd1,GGMd} is employed,
commonly referred to as the ``Snowmass Points and Slopes 8'' (SPS8) benchmark model
\cite{Allanach:2002nj}.
In this scenario, pair-produced squarks and gluinos undergo cascade decays as
shown in Fig.~\ref{fig:feyn}, and eventually produce the lightest SUSY
particle (LSP), the gravitino (\PXXSG), which is stable and weakly interacting.
The phenomenology of such decay chains is primarily determined by the nature of the next-to-lightest
SUSY particle (NLSP). In the SPS8 benchmark, the NLSP is the
lightest neutralino, \PSGczDo, and the mass of the NLSP is linearly related to the
effective scale of SUSY breaking, $\Lambda$~\cite{Giudice:1998bp,Chen:1997yfa}.

In the SPS8 model, $\Lambda$ is a free parameter whose value determines the primary production mode
and decay rate of SUSY particles. Depending on the value of $\Lambda$,
the coupling of the NLSP to the gravitino could be very weak and lead to long NLSP lifetimes.
The dominant decay mode of the NLSP is to a photon and a gravitino, resulting in a final state
with one or two photons and missing transverse momentum (\ptmiss).
The dominant squark-pair and gluino-pair production modes also result in additional energetic jets.
If the NLSP has a proper decay length that is a significant fraction of the
radius of the CMS tracking volume (about 1.2\m),
then the photons produced at the secondary vertex tend to exhibit distinctive features.
Because of their production at displaced vertices and their resulting trajectories, the photons
have significantly delayed arrival times (order of ns) at the CMS electromagnetic calorimeter (ECAL) compared to particles produced
at the primary vertex and traveling at the speed of light.
They also enter the ECAL at non-normal impact angles.

The present search makes use of these features to
identify potential signals of physics beyond the SM. We select events
with one or two displaced or delayed photons, and
three or more jets. Signal events are expected to produce large \ptmiss
as the LSP escapes the detector volume without detection. In the case of very long-lived NLSPs, one of the NLSPs
may completely escape the detector, further increasing the \ptmiss.
Previously, similar searches for LLPs decaying to displaced or delayed photons have been performed by
the CMS~\cite{Chatrchyan:2012jwg} and ATLAS~\cite{Aad:2014gfa} Collaborations using
LHC collisions at a center-of-mass energies of 7~and~8\TeV, respectively.
Past LHC searches for invisible Higgs boson decays in association with
photons~\cite{Khachatryan:2015vta} also have sensitivity to such models.

\section{The CMS detector}
\label{sec:detector}

The central feature of the CMS apparatus is a superconducting solenoid of 6\m internal diameter,
providing a magnetic field of 3.8\unit{T}. Within the solenoid volume are a silicon pixel and strip
tracker, a lead tungstate crystal ECAL, and a brass and scintillator hadron calorimeter (HCAL), each
composed of a barrel and two endcap sections. Forward calorimeters extend the pseudorapidity
coverage provided by the barrel and endcap detectors. Muons are measured in gas-ionization
detectors embedded in the steel flux-return yoke outside the solenoid.

The ECAL is highly granular and consists of 61\,200 crystals in the barrel region, each with an area of
approximately $2.2{\times}2.2\cm^{2}$ corresponding to roughly $0.0174{\times}0.0174$ in
$\eta$-$\phi$ space, where $\eta$ is the pseudorapidity and $\phi$ the azimuthal angle (in radians)
of the coordinate system~\cite{Chatrchyan:2008zzk}.
Each of the two endcap sections consist of 7324 crystals, each crystal having an area of
$2.68{\times}2.68\cm^{2}$.
A typical electromagnetic shower spans approximately 10 crystals with energy deposits
above noise threshold.
The barrel and endcap ECAL components cover the regions with $\abs{\eta} < 1.5$ and
$1.5 < \abs{\eta} < 2.5$,
respectively. The best possible time resolution for each ECAL channel is
measured to be between 70 and 100\ps, depending on detector aging.

The first level of the CMS trigger system~\cite{Khachatryan:2016bia}, composed of custom hardware processors, uses information from the
calorimeters and muon detectors to select the most interesting events in a fixed time interval of less than 4\mus. The high-level
 trigger (HLT) processor farm further decreases the event rate from around 100\kHz to less than 1\kHz, before data storage.
A more detailed description of the CMS detector, together with a definition of the coordinate system used and the relevant
kinematic variables, can be found in Ref.~\cite{Chatrchyan:2008zzk}.

\section{Event samples}
\label{sec:samples}

This analysis uses data sets of proton-proton ($\Pp\Pp$) collisions collected by the
CMS experiment at the LHC in 2016 and 2017, corresponding to integrated
luminosities of 35.9 and 41.5\fbinv, respectively.
Simulated samples are used to study the SM background and signal contributions,
primarily for the purpose of optimizing the event selection and the binning in the photon time
and \ptmiss observables. The \MGvATNLO v2.2.2 generator~\cite{Alwall:2007fs} is used at
next-to-leading order (NLO) in quantum chromodynamics (QCD) to simulate events originating from
single top quark and top quark pair production,
and at leading order (LO) to simulate
events originating from QCD multijet, \gjets, \wjets, and \zjets production.
Simulated samples of diphoton events are generated using \SHERPA~v2.2.4~\cite{Bothmann:2019yzt,Gleisberg:2008ta},
and include Born processes with up to three additional jets, as well as box processes at LO precision.
The particle spectra of each GMSB SPS8 signal model are tabulated in a SUSY Les Houches accord (SLHA) file
using \ISASUGRA as part of \ISAJET v7.87~\cite{Paige:2003mg}. The SLHA files are then used to generate benchmark signal model
samples using \PYTHIA~v8.212~(v8.230)~\cite{Sjostrand:2014zea} for the 2016 (2017) data analysis.

For all simulated samples discussed above, the fragmentation and parton showering are modeled
using \PYTHIA v8.212 with the CUETP8M1 underlying event tune~\cite{Skands2014,CMS-PAS-GEN-14-001}
(\PYTHIA v8.230 with the CP5~\cite{CMS-PAS-GEN-17-001} tune) for the 2016 (2017)
data analysis. The {NNPDF3.0}~\cite{Ball:2014uwa}~and~{NNPDF3.1}~\cite{Ball:2017nwa} parton distribution
function (PDF) sets are used for the 2016 and 2017 simulated samples, respectively.
The signal and background samples are processed through a simulation
of the CMS detector based on \GEANTfour~\cite{geant4} and are reconstructed with the same
algorithms as used for data. Additional $\Pp\Pp$ interactions in the same or
adjacent bunch crossings, referred to as pileup, are also simulated.

\section{Trigger and event selection}
\label{sec:selection}
The unique signature of delayed photons is best exploited with
specialized triggers and dedicated photon reconstruction and identification criteria.
There is a difference between the search selections for the 2016 and 2017
data sets, primarily because of the introduction of a targeted HLT
algorithm implemented for the 2017 data set, which superseded a
general diphoton trigger used for the 2016 data set.

\subsection{Trigger selection}
\label{sec:trigger}

For the 2016 data set, events are selected by the standard diphoton trigger,
requiring transverse momenta (\pt) larger than 42~and~25\GeV for the leading
and subleading photons, respectively. Loose identification criteria
are imposed on the photon shower width in the ECAL and
on the ratio of the energies recorded in the ECAL and HCAL
to reduce the rate of background from jets misidentified as photons.

For the 2017 data set, a dedicated HLT algorithm was developed to select
events with a single photon satisfying requirements consistent with production at a displaced vertex.
Such photons tend to strike the
front face of the barrel ECAL at a non-normal incidence angle, resulting
in a more elliptical electromagnetic shower in
the $\eta$-$\phi$ plane~\cite{Chatrchyan:2012jwg}.
In addition to standard requirements on the shower width and
electromagnetic to hadronic energy ratio, requirements on the
major and minor axes of the shower are also imposed. This allows the identification of the
elliptical shower shape, described in greater detail in Sec.~\ref{sec:objects}.
Loose requirements on the amount of energy around the direction of the photon in the CMS subdetectors (isolation)
are also imposed on trigger photon candidates,
and the photon \pt is required to exceed 60\GeV.
Electrons misidentified as photons are suppressed by requiring the candidate photon
to be geometrically isolated from charged-particle tracks.
Relaxing the trigger requirement from two photons to only one photon
increases the background rate, and in order to reduce the trigger rate
to a level acceptable for the operation of the HLT the scalar \pt
sum of all jets (\HT) is required to exceed 350\GeV. For signals with
neutralino proper decay length larger than 10\m, the signal
acceptance is improved by about a factor of two compared to the 2016 data set.

\subsection{Object reconstruction and selection}
\label{sec:objects}
A particle-flow (PF) algorithm~\cite{Sirunyan:2017ulk} is used to
reconstruct and identify each individual particle in an event
using an optimized combination of information from the
various elements of the CMS detector.
The candidate vertex with the largest value of summed physics-object $\pt^2$ is
taken to be the primary $\Pp\Pp$ interaction vertex. The physics objects are the jets,
clustered using the jet finding algorithm~\cite{Cacciari:2008gp,Cacciari:2011ma}
with the tracks assigned to candidate vertices as inputs, and the associated missing
transverse momentum, taken as the negative vector sum of the \pt of those jets.

Photon candidates are reconstructed from energy clusters in the
ECAL~\cite{Khachatryan:2015iwa} and identified
based on the transverse shower width, the hadronic to electromagnetic
energy ratio, and the degree of isolation from charged particle tracks.
Photons are required to satisfy $\abs{\eta} < 2.5$ and to not fall
in the transition region between the barrel and endcap of the
ECAL ($1.444 < \abs{\eta} < 1.566$), where the photon reconstruction
is not optimal. For the 2016 data set,
photon candidates that share the same energy cluster as an identified
electron associated with the primary vertex
are vetoed following the procedure detailed in Ref.~\cite{Khachatryan:2015iwa}.
To remain consistent with the HLT selection, photons
matched geometrically to charged-particle tracks are vetoed for the 2017 data set as well.

Because of algorithms designed to reject noise and out-of-time pileup, the default
photon reconstruction vetoes photons delayed by more than 3\ns.
To evade this veto, a second set of out-of-time (OOT) photons is therefore defined,
in which the clustering starts from ECAL deposits whose signals are delayed by more than 3\ns.
The remainder of the reconstruction algorithm for OOT photons is identical to
the standard photon reconstruction described in the previous paragraph.
In addition to being delayed, signal photons tend to impact the
front face of the barrel ECAL at a non-normal incidence angle, and yield
electromagnetic showers that are more elliptical in the $\eta$-$\phi$ plane.
To make use of this discriminating feature, we define the OOT photon
identification criteria  including selection requirements on the \smaj and \smin observables
defined as:
\begin{equation}
\begin{aligned}
\label{eqn:SminorSmajor}
\smaj & =\frac{\spp + \see + \sqrt{(\spp - \see)^2 + 4\sep^2}}{2}, \\
\smin & =\frac{\spp + \see - \sqrt{(\spp - \see)^2 + 4\sep^2}}{2}
\end{aligned}
\end{equation}
where \spp, \see, and \sep are the second central moments of the spatial distribution
of the energy deposits in the ECAL in $\eta$-$\phi$ coordinates, and are proportional to the
squared lengths of the semimajor and semiminor axes of the elliptical shower shape.
The full set of criteria for the OOT photon selection additionally includes requirements on
the transverse shower width   and isolation and was obtained through a separate optimization
that maximizes the discrimination between displaced signal photons and background photons associated
with the primary vertex.

Hadronic jets are reconstructed by clustering PF candidates using
the anti-\kt algorithm with a distance parameter of 0.4~\cite{Cacciari:2008gp, Cacciari:2011ma}.
Further details of the performance of the jet reconstruction can
be found in Ref.~\cite{CMS-PAS-JME-16-003}.  Jets used in any selection of
this analysis are required to have $\pt > 30\GeV$ and $\abs{\eta} < 3.0$.

The negative vector \pt sum of all the PF candidates in an event
is defined as \ptvecmiss, and its magnitude is denoted as
\ptmiss~\cite{Sirunyan:2019kia}. The \ptvecmiss
is modified to account for corrections to the energy scale of the reconstructed
jets in the event. Because OOT photons are not part of the standard PF candidate
reconstruction used to compute the \ptvecmiss, we correct the \ptvecmiss by
adding the negative momentum of an OOT photon if it is selected in the event.
Anomalous high-\ptmiss events can arise because of a variety of reconstruction failures,
detector malfunctions, or noncollision backgrounds.
Filters for vetoing such anomalous events are applied~\cite{Sirunyan:2019kia}.

\subsection{Photon time reconstruction}
\label{sec:photontime}

Photons from signal events tend to arrive at the ECAL up to 10\ns later than
particles produced at the primary vertex. Therefore measuring the photon time of
arrival delay with respect to a photon produced
at the primary vertex and traveling at the speed of light helps to discriminate between
signal and background. The time of arrival of a photon at the ECAL, \tECAL, is calculated based on a
weighted sum of the arrival times reconstructed from the signal pulse in each
ECAL crystal comprising the photon cluster:
\begin{equation}
\label{eqn:timestamp}
\tECAL = \frac{\sum_{i}{\frac{\tECALi}{\sigSQi}}}{\sum_{i}{\frac{1}{\sigSQi}}},
\end{equation}
where \tECALi is the timestamp of the signal pulse in crystal $i$~\cite{Chatrchyan:2009aj}.
The estimated time resolution of the signal pulse in crystal $i$ is $\sigma_{i}$ and is parametrized as:
\begin{equation}
\label{eqn:timeResolution}
\sigma^{2}_{i} = \left(\frac{N}{\Ai/\sigNi}\right)^{2} + C^{2},
\end{equation}
where \Ai is the amplitude of the signal detected by crystal $i$, \sigNi is the
pedestal noise for crystal $i$, and $N$ and $C$ are constants fitted from a dedicated
measurement of the time resolution of the crystal sensors.

\begin{figure}[hbtp] \centering
\includegraphics[width=1.0\columnwidth]{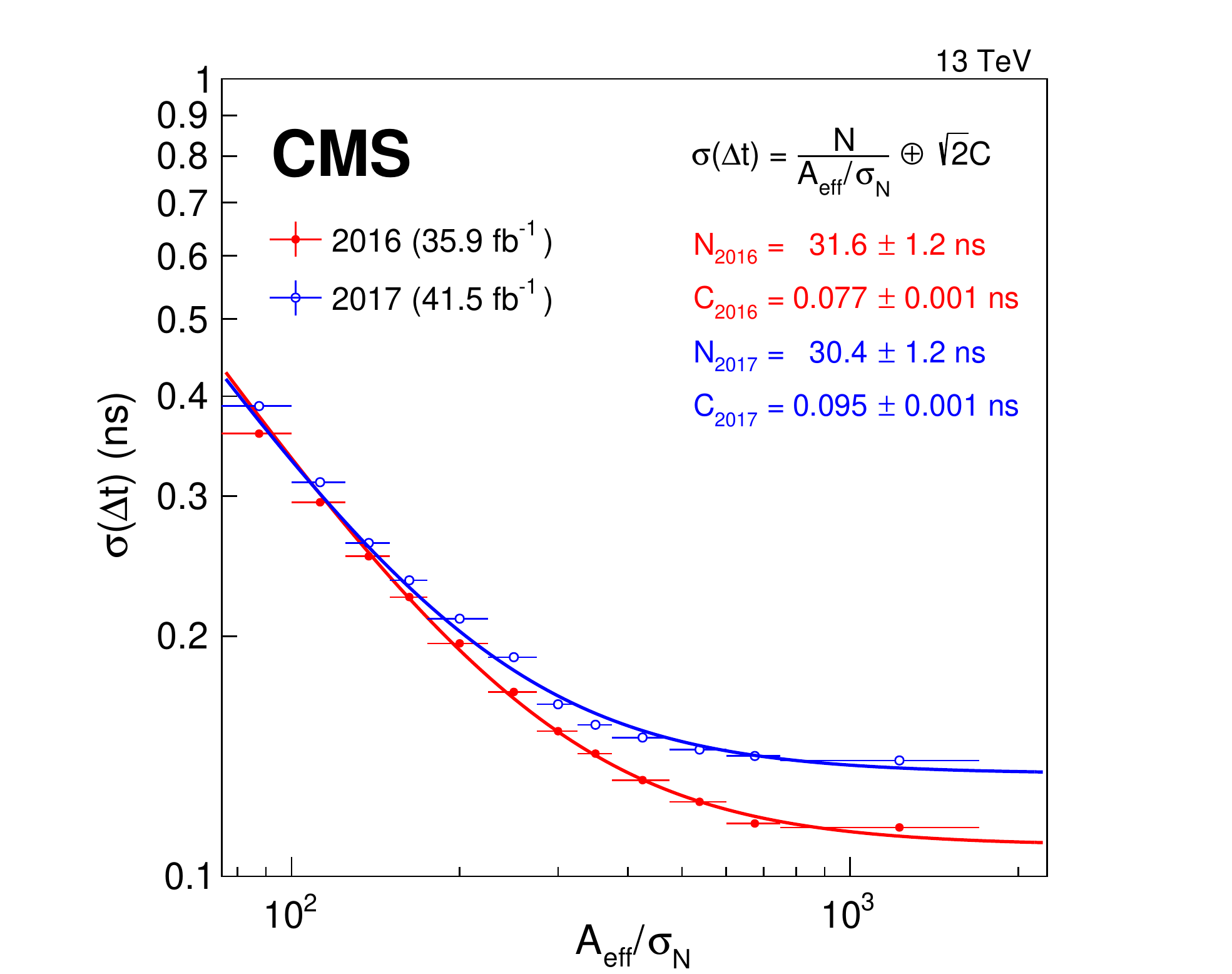}
\caption{ The time resolution between two neighboring ECAL crystals as a function of the
effective amplitudes of the signals in the two crystals for the 2016 and
2017 data sets. The lines shown reflect the fits described in the text. The horizontal bars
on the data represent the bin widths, which are treated as uncertainties in the fit.}
\label{fig:ECALTimeResolution}
\end{figure}

To measure the crystal sensor time resolution, we follow a procedure similar to that described in
Refs.~\cite{Chatrchyan:2009aj,delRe:2015hla}. We first apply a very loose selection on
photons using \smaj and \smin in order to reject jets.
Pairs of crystals from the same photon cluster are selected by requiring
that their energies are within $20\%$ of each other, are nearest neighbors either in the $\eta$ or $\phi$
directions, and are within the same $5{\times}5$ grid of crystals defining a trigger tower.
The distributions of time differences measured in such crystal pairs are fitted using
Gaussian functions in bins of
the effective amplitude \AeosN, and the standard deviation of each fitted Gaussian
function is trended as a function of \AeosN. The effective amplitude is
obtained combining the signals in the two crystals and is denoted by:
\begin{equation}
\label{eqn:effAmp}
\AeosN = \frac{\AsigNX{1}\AsigNX{2}}{\sqrt{\AsigNX{1}^2+\AsigNX{2}^2}}.
\end{equation}
The results for the 2016 and 2017 data sets are shown in Fig.~\ref{fig:ECALTimeResolution}.
These resolution measurements are fitted with the functional form given by
Eq.~(\ref{eqn:timeResolution}), and the $N$ and $C$ parameters are extracted and
summarized in Table~\ref{tab:EcalTimingParameters}.
These parameters are then used to calculate the weights for the photon timestamp in Eq.~(\ref{eqn:timestamp}).
The observed worsening of the constant term to the time resolution in 2017 may be due to a progressive
loss of transparency of the crystals from radiation damage.

\begin{table}[htb!]
\centering
\topcaption{The fitted ECAL timing resolution parameters for the 2016 and 2017 data sets.}
\label{tab:EcalTimingParameters}
\begin{scotch}{ l c c }
Parameters & 2016 Data set & 2017 Data set \\
\hline
N & $31.6 \pm 1.2$\ns\phantom{0} & $30.4 \pm 1.2$\ns\phantom{0} \\
C & $0.077 \pm 0.001$\ns & $0.095 \pm 0.001$\ns\\
\end{scotch}
\end{table}

To calibrate the photon timestamp response, electrons from \Zepem decays with an invariant mass
between $60$~and~$150$\GeV are reconstructed as photons.  For each such photon candidate, the \tECAL is
adjusted for the time-of-flight between the primary vertex
and the location of the impact of the photon on the front
face of ECAL. The timestamp for each photon is recorded, and the mean and RMS parameters of the
resulting distribution are extracted as a function of the photon energy. The time response mean
is adjusted to zero for both data and simulation,
and the timestamps in the simulated events are smeared by an additional Gaussian-distributed random variable
such that the resolution in simulation matches that measured in data. The
calibrated photon arrival time is denoted as \tpho. These calibrations are
applied to simulated signal samples
in order to accurately predict the signal response, and their uncertainties
are propagated to the predicted shape  of the \tpho distribution for the
signal as a systematic uncertainty. The time resolution of a single
photon candidate is roughly 400\ps. The resolution is constant up to a photon timestamp of
25\ns, the upper boundary of \tpho used during the signal extraction.

\subsection{Event selection}
\label{sec:eventSelection}

Events with at least one photon in the barrel region of the detector ($\abs{\eta} < 1.444$)
with \pt larger than 70\GeV are selected. Standard photons~\cite{Khachatryan:2015iwa}
and OOT photons are required to pass the ``tight'' working points. Both photon
identifications are tuned to have an average efficiency of about $70\%$.
Furthermore, a displaced photon identification requirement
based on the \smaj and \smin variables is imposed.
The calibrated arrival time of this tight photon, \tpho, is used as one of the final discriminating observables
to distinguish signal from background.
For the dominant squark-pair and gluino-pair production modes shown in Fig.~\ref{fig:feyn},
the NLSP is generally produced in association with several jets, and therefore
we also require events to have three or more jets with \pt larger than 30\GeV.

In order to remain compatible with the respective HLT selection,
slightly different event selection criteria are imposed
on the 2016 and 2017 data sets. For the 2016 data set, triggered by a diphoton HLT,
a second photon with \pt larger than 40\GeV is required to match the analogous HLT requirement.
For the 2017 data set, the first category, referred to as the \onepho{2017} category, requires events with no
subleading photon or events where the subleading photon
does not pass the photon identification criteria. The second category requires events to
have a subleading photon satisfying the photon identification criteria, and is referred
to as the \twopho{2017} category.
The second-photon requirement helps to reduce background by one to two orders of magnitude,
while the signal yield remains high for low to intermediate lifetimes.
Finally, for the 2017 data set, the \HT is required to be larger
than 400\GeV in order to match the requirements of the
HLT and to reach the plateau of the trigger efficiency.

For the 2016 and \twopho{2017} analyses, for a given neutralino proper decay length, the signal yield
increases as a function of the SUSY breaking scale, $\Lambda$, by roughly a factor of two over the range
considered for this analysis ($\Lambda$~from 100 to 400\TeV).
The product of signal efficiency and acceptance for the lowest $\Lambda$ is roughly $10.0\pm0.1\%$ and $0.15\pm0.01\%$
for neutralino proper decay lengths of 0.1 and 100\m, respectively. For the \onepho{2017} analysis, the product of
signal efficiency and acceptance varies as a function of $\Lambda$ from $5.5\pm0.1$ to $10.4\pm0.2\%$ for a neutralino proper decay length of 0.1\m,
and from $0.22\pm0.03$ to $0.65\pm0.05\%$ for a neutralino proper decay length of 100\m.
These trends can be explained by the harder photon spectrum and increase in jet
activity that result from an increase in $\Lambda$, while an
increase in the neutralino proper decay length results in either
one or both of the NLSPs decaying outside the fiducial region of ECAL.

\begin{figure*}[hbtp]
\centering
\includegraphics[width=0.490\textwidth]{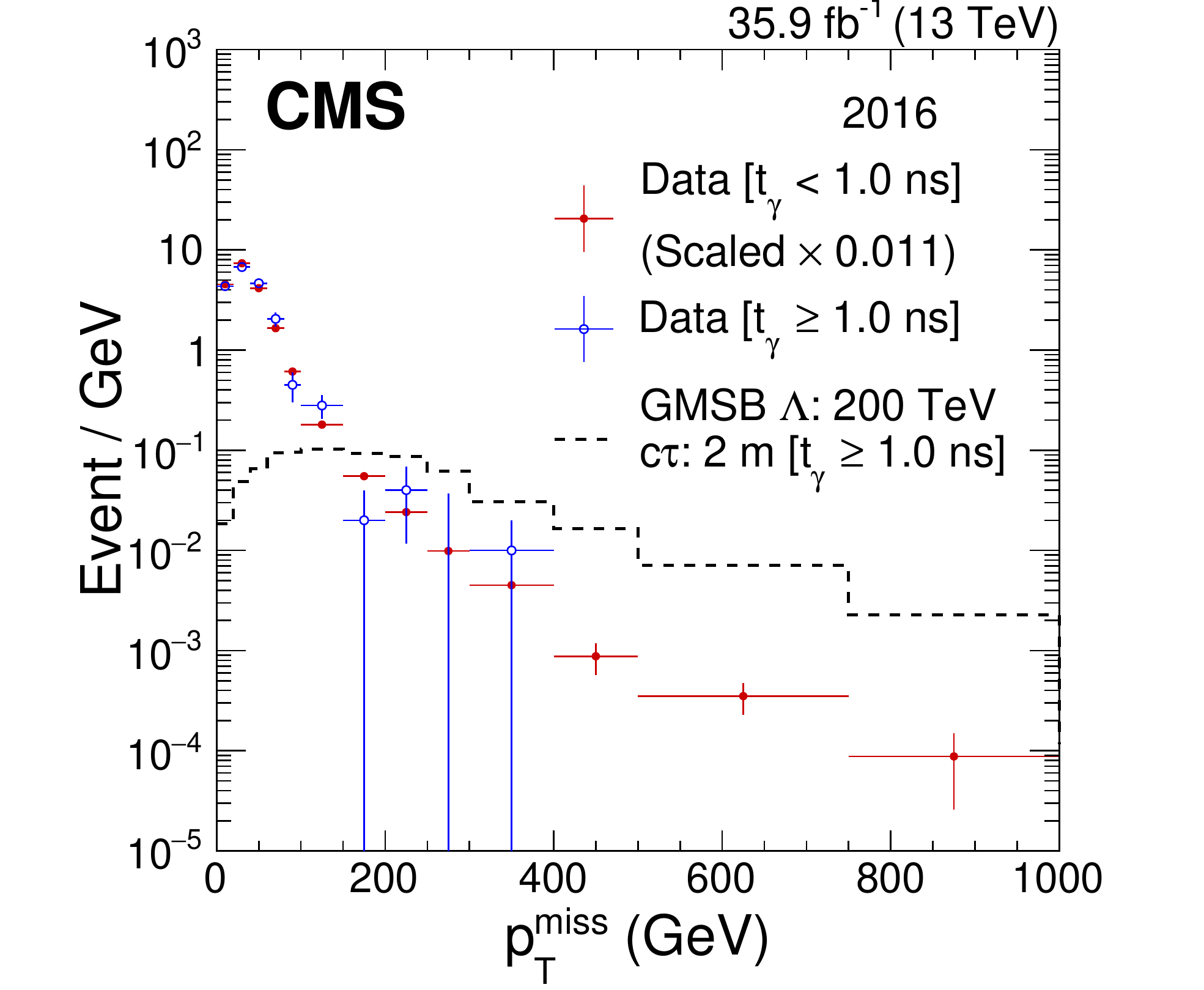}
\includegraphics[width=0.490\textwidth]{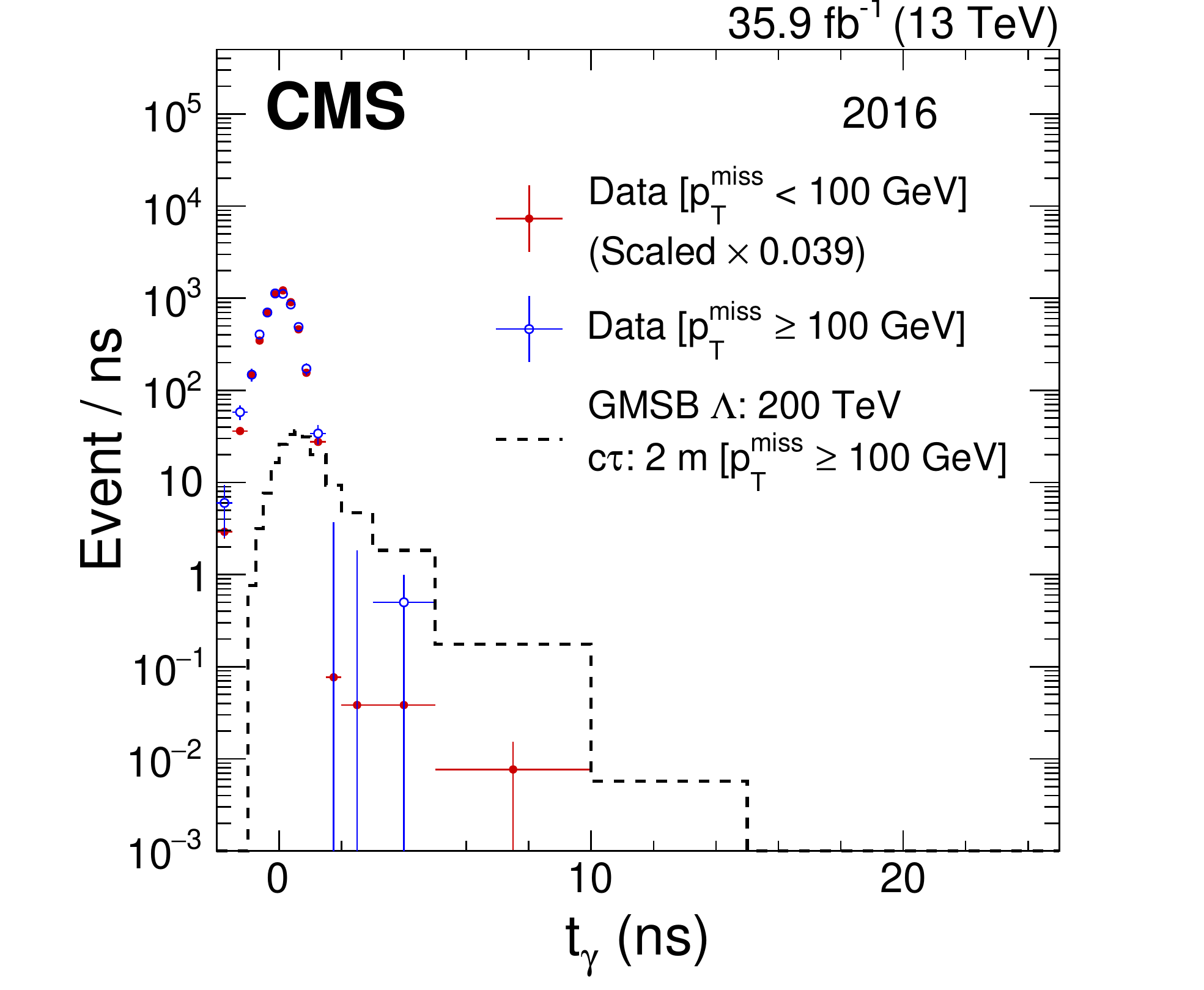}

\caption{The \ptmiss (left) and \tpho (right) distributions for the 2016 event selection,
shown for data and a representative signal benchmark (GMSB: $\Lambda=200\TeV, \ctau=2\m$).
The \ptmiss distribution for data is separated into events with
$\tpho\geq1\ns$ (blue, darker) and $\tpho<1\ns$ (red, lighter), scaled to match the
total number of events with $\tpho\geq1\ns$.
The \tpho distribution for data is separated into events with
$\ptmiss\geq100\GeV$ (blue, darker) and $\ptmiss<100\GeV$ (red, lighter),
scaled to match the total number of events with $\ptmiss\geq 100\GeV$.
The signal (black, dotted) is shown in the left plot only for events with
$\tpho\geq1\ns$, and in the right plot only for events with $\ptmiss\geq 100\GeV$.
The entries in each bin are normalized by the bin width.
The horizontal bars on data indicate the bin boundaries.
The last bin in each plot includes overflow events.}
\label{fig:time_met_1D_2016}
\end{figure*}

\begin{figure*}[hbtp]
\centering
\includegraphics[width=0.490\textwidth]{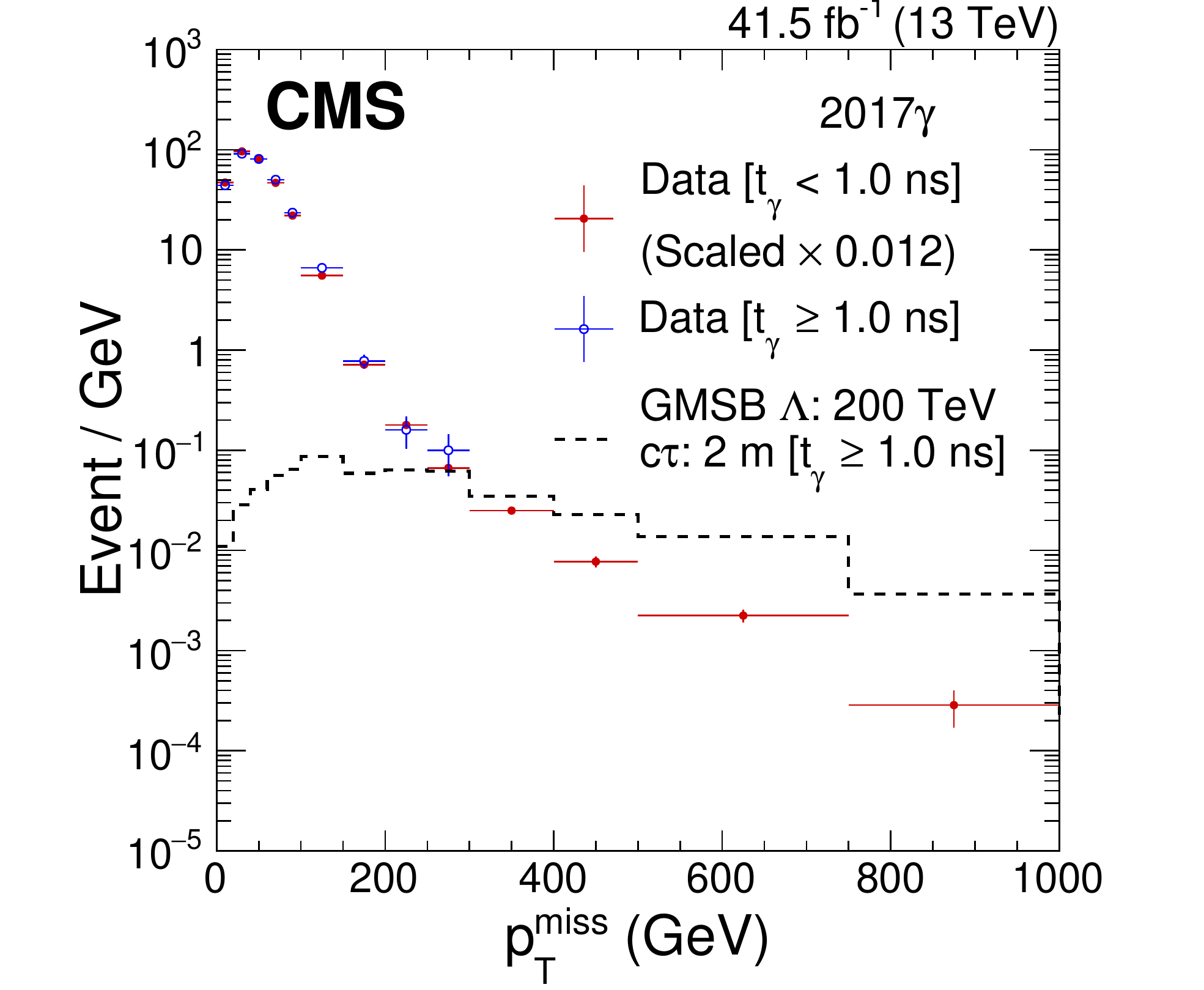}
\includegraphics[width=0.490\textwidth]{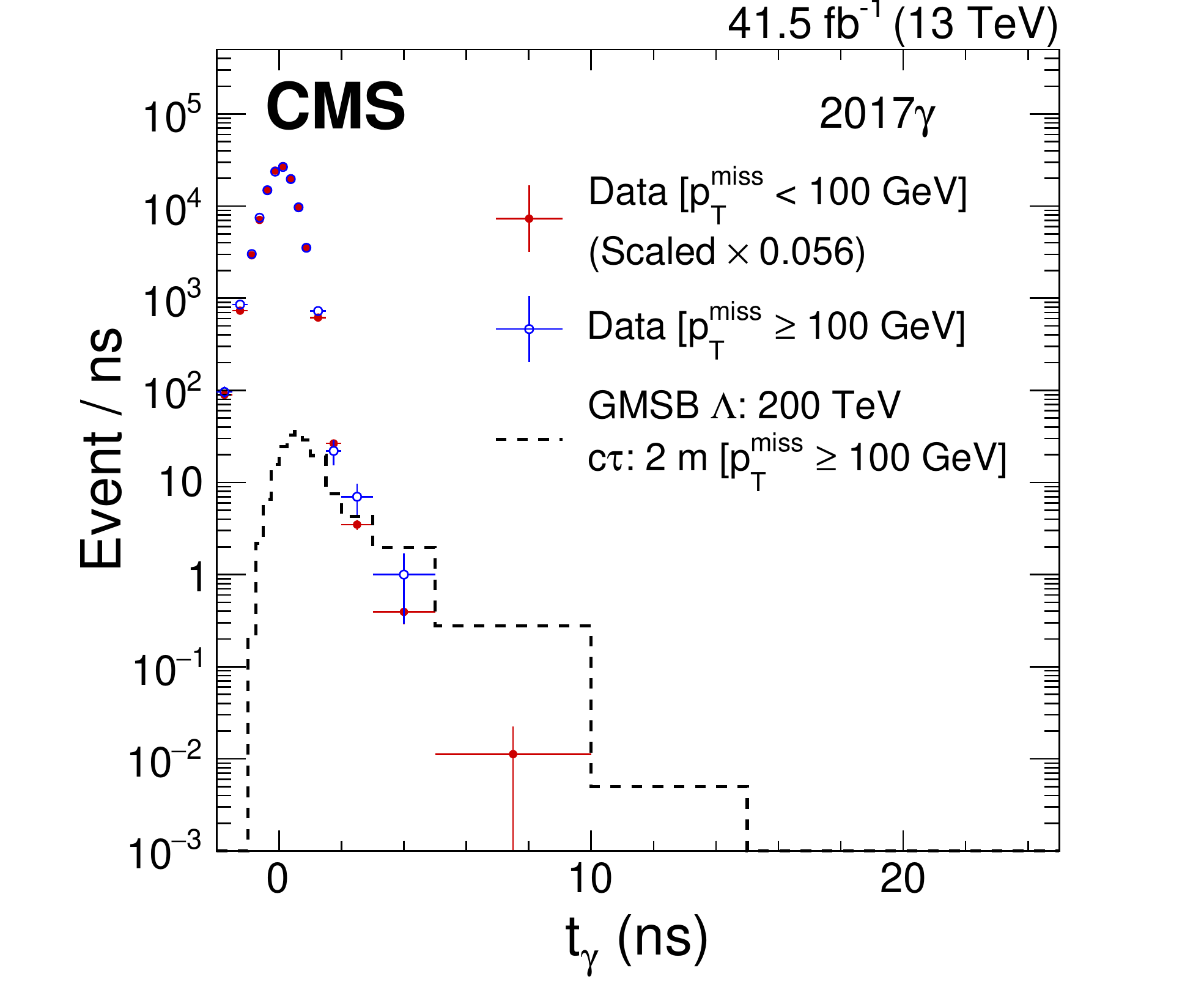}
\includegraphics[width=0.490\textwidth]{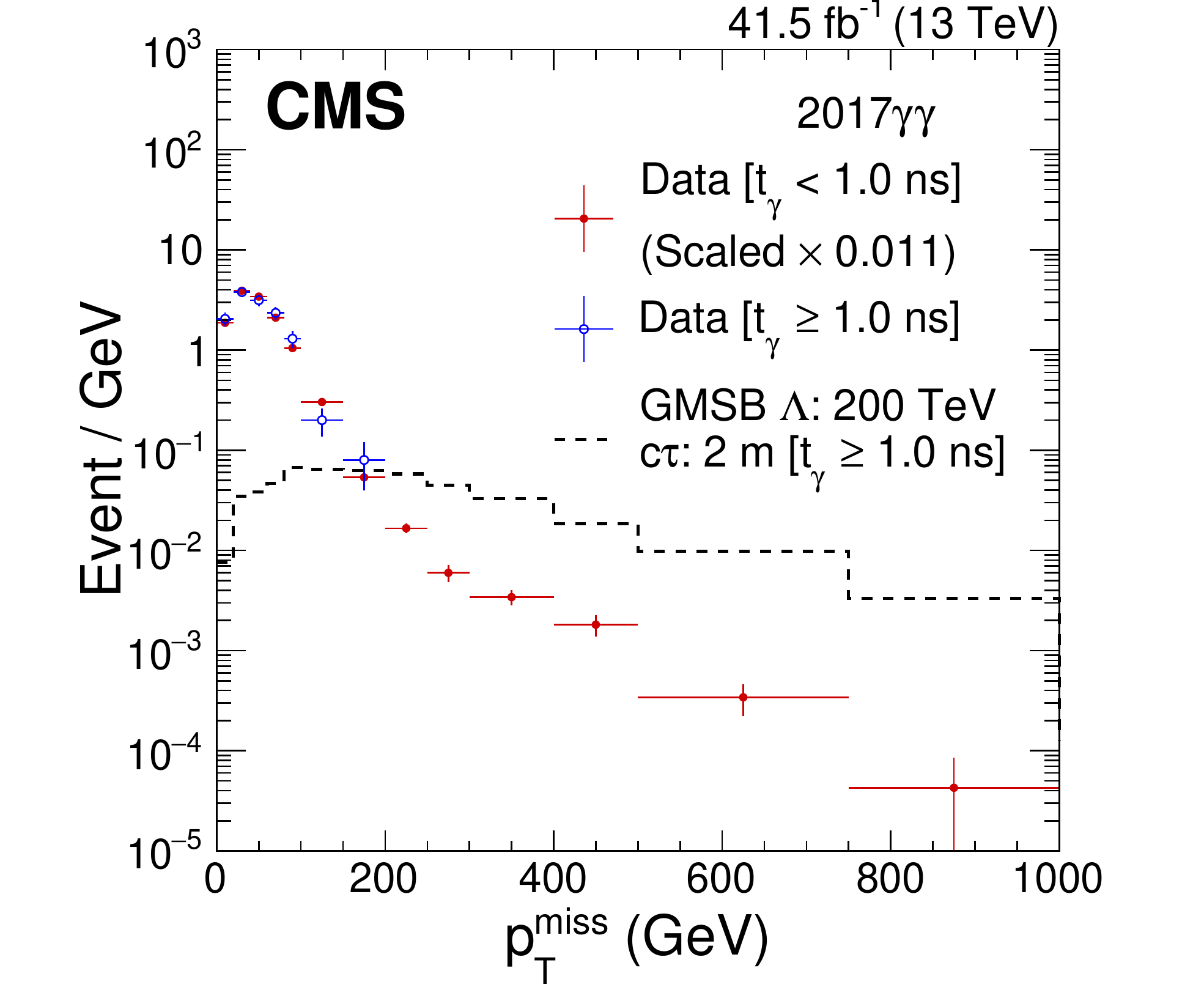}
\includegraphics[width=0.490\textwidth]{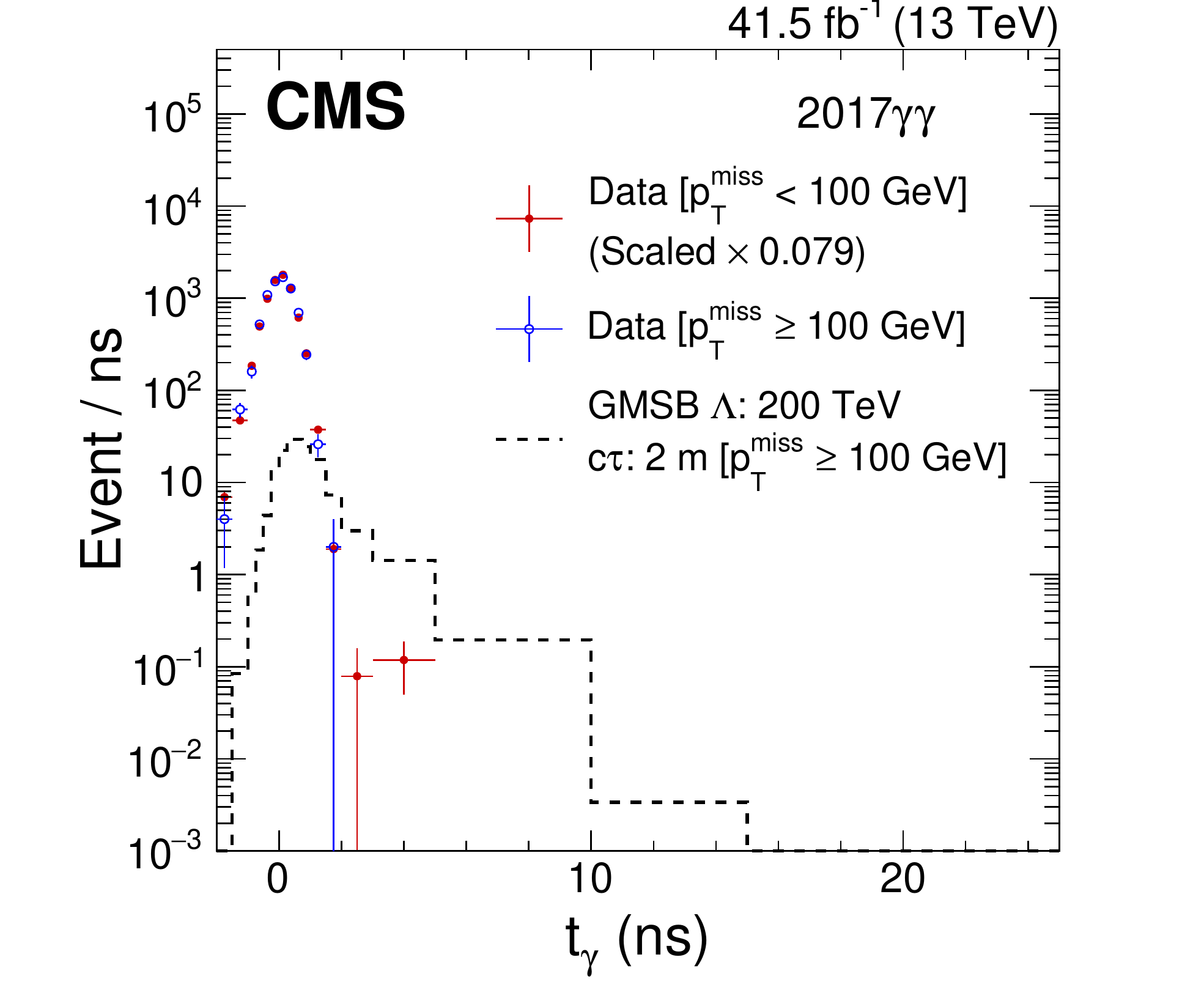}

\caption{ The \ptmiss (left) and \tpho (right) distributions
for the \onepho{2017} (upper row) and \twopho{2017} (lower row)
event selections shown for data and a representative
signal benchmark (GMSB: $\Lambda=200\TeV, \ctau=2\m$).
The \ptmiss distribution for data is separated into events with
$\tpho\geq1\ns$ (blue, darker) and $\tpho<1\ns$ (red, lighter), scaled to match the
total number of events with $\tpho\geq1\ns$.
The \tpho distribution for data is separated into events with
$\ptmiss\geq100\GeV$ (blue, darker) and $\ptmiss<100\GeV$ (red, lighter),
scaled to match the total number of events with $\ptmiss\geq 100\GeV$.
The signal (black, dotted) is shown in the left plots only for events with
$\tpho\geq1\ns$, and in the right plots only for events with $\ptmiss\geq 100\GeV$.
The entries in each bin are normalized by the bin width.
The horizontal bars on data indicate the bin boundaries.
The last bin in each plot includes overflow events.}
\label{fig:time_met_1D_2017}
\end{figure*}

Figures~\ref{fig:time_met_1D_2016}~and~\ref{fig:time_met_1D_2017}
show the \ptmiss (\tpho) distribution in data
for low and high \tpho (low and high \ptmiss), for the 2016, \onepho{2017}, and \twopho{2017} event
selections. In addition, the distribution of events for a representative signal point (GMSB:
$\Lambda=200\TeV, \ctau=2\m$) is also shown, scaled by the product of the production cross section and the
integrated luminosity in the regions most sensitive to this signal benchmark: large \ptmiss and \tpho.

\section{Signal extraction and background estimation}
\label{sec:bkg}

The \ptmiss and \tpho variables are used as the final discriminating observables
to distinguish signal from background. Standard model background events can populate the
signal-enriched regions with large values of \ptmiss and \tpho because of imperfect
resolution. Four bins are defined based on the values
of the \ptmiss and \tpho observables.
Bin A has low \ptmiss and low \tpho; bin B has high \ptmiss and low \tpho; bin C
has high \ptmiss and high \tpho; and bin D has low \ptmiss and high \tpho.
Signals with large lifetimes are concentrated in bin C, while signals with shorter lifetimes
tend to occupy bin B. In contrast, backgrounds are concentrated in bin A.
In general, bin C is the most sensitive, with largest signal to background ratio.
After the offline selection is applied, the main background contribution is from $\Pp\Pp$ collision processes
with high \ptmiss, which have the same timing distribution as low-\ptmiss collider data, ensuring
that the two discriminating variables are independent for background processes.
This includes proton collisions from satellite bunches spaced $\sim2.5\ns$ apart from the main bunches.
The noncollision backgrounds, which include cosmic ray muons, beam halo muons, and electronic noise
deposits, are reduced to a negligible level by the jet multiplicity requirement and the photon selections.

As the \ptmiss and \tpho observables are statistically independent for background processes,
the background distribution can be factorized into the product of the distributions of these two observables.
This permits the use of the so called ``ABCD'' method to predict the background yield
in the signal-enriched bin C as $\NX{C} = (\NX{D}\NX{B})/\NX{A}$,
where $\NX{X}$ is the number of background events.
In order to account for potential signal contamination in bins A,~B,~and~D,
a modified ABCD method is used where a binned maximum likelihood fit is performed
simultaneously in the four bins, with the signal strength
included as a floating parameter that scales the signal yield uniformly in each bin.
The background component of the fit is constrained to obey the
standard ABCD relationship, within the bounds of a small systematic uncertainty
derived from a validation check of the method in a control region (CR).
Systematic uncertainties that impact the signal and background yields are
treated as nuisance parameters with log-normal probability density functions.

For each point in the signal model parameter
space ($\Lambda$ and \ctau in Table~\ref{tab:optimizedBinning}), the
boundaries in \ptmiss and \tpho that define the A, B, C, and D bins
are chosen to yield optimal expected sensitivity.
For the optimization procedure, in order to remain unbiased by the observed data in the signal-enriched regions,
we estimate the background yields using only the observed yield in data for
bin A (\NX{A}) as follows.
Template shapes for the observable \ptmiss (\tpho) are derived from data
requiring that $\abs{\tpho}<1\ns$ ($\ptmiss < 100\GeV$). These regions are
defined to have negligible signal yield.
We obtain the ratios $\rXA{B}$ ( $\rXA{D}$ ) by dividing the number of events
with \ptmiss ( \abs{\tpho} ) larger than the given bin boundary by the number of events
with \ptmiss ( \abs{\tpho} ) smaller than the bin boundary. The background yields in bins B, D, and C are
calculated as $\NX{A}\rXA{B}$, $\NX{A}\rXA{D}$,
and $\NX{A}\rXA{B}\rXA{D}$, respectively.
The resulting optimized bin boundaries in \tpho and \ptmiss are obtained by choosing
the bin boundaries that yield the best expected limit and are summarized
in Table~\ref{tab:optimizedBinning} for all the SPS8 model parameter space points considered.
To simplify the analysis, groups of similar signal model parameters share the same optimized bin boundaries.

It should be noted that we set the lower and upper boundaries in \tpho to be -2\ns and 25\ns,
respectively. The lower boundary is set by five times the single photon candidate time resolution, while
the upper boundary is set to avoid contamination from the next LHC bunch crossing.

\begin{table*}[htb!]
\centering
\topcaption{The optimized bin boundaries for \tpho (first number, in units of ns)
and \ptmiss (second number, in units of \GeV), for different GMSB SPS8
signal model benchmark points considered in the search and for each data set
category.}
\label{tab:optimizedBinning}
\begin{scotch}{l c c c c c c }
\multirow{2}{*}{\ctau (m)} & \multicolumn{3}{c} {$\Lambda \leq 300 \TeV$} & \multicolumn{3}{c} {$\Lambda> 300 \TeV$} \\
 & 2016 & \onepho{2017} & \twopho{2017} & 2016 & \onepho{2017} & \twopho{2017} \\
\hline
(0, 0.1) & 0\hphantom{.5} , 250 & 0.5 , 300 & 0.5 , 150 & 0\hphantom{.5} , 250 & 0.5 , 300 & 0.5 , 200 \\
(0.1 ,  100) & 1.5 , 100 & 1.5 , 200 & 1.5 , 150 & 1.5 , 150 & 1.5 , 300 & 1.5 , 200 \\
\end{scotch}
\end{table*}

To verify that the \ptmiss and \tpho observables are independent, we
define CRs that isolate different SM processes that are similar to the
backgrounds expected in the signal region (SR). The \gjets CR,
dominated by the \gjets process, is defined
as events satisfying the same requirements as the SR, but having
fewer than three jets. The multijet CR, dominated by QCD multijet production,
comprises events satisfying the same requirements as the SR,
but with an inverted isolation requirement on the leading photon.
We measure the correlation coefficients between \ptmiss and \tpho to be less
than 1\% for both the \gjets CR and multijet CR, supporting their independence.
A closure test on the predicted background yield in these CRs
is propagated as a systematic uncertainty, as discussed further in
Sec.~\ref{sec:uncertainties}.

\section{Systematic uncertainties}
\label{sec:uncertainties}

The dominant uncertainty in the search is the statistical uncertainty
in the background prediction of the modified ABCD method.
There are several subdominant systematic uncertainties that affect the prediction of the
signal yield in all four bins. These systematic uncertainties include the
uncertainty in the integrated luminosity measurement~\cite{CMS-PAS-LUM-17-001,CMS-PAS-LUM-17-004},
in the energy scale and resolution of the photons and jets,
and in the trigger and photon identification efficiencies.
For all these cases, dedicated measurements are performed that evaluate corrections
and uncertainties in the efficiencies and energy scales in simulated signal events,
and these uncertainties are propagated to the signal yield predictions as an uncertainty
in the predicted shapes of the distributions of the discriminating observables \ptmiss and \tpho. The calibration
of the timestamp discussed in Sec.~\ref{sec:photontime} has associated uncertainties
that affect both the offset and the resolution in \tpho, and are propagated
in the shape prediction for the \tpho distribution
for the signal benchmarks. As we use \Zepem events to measure the photon identification efficiency,
the corresponding systematic uncertainty includes the impact of the difference in detector
response between an electron and a photon. Table~\ref{tab:syst_summary} provides a summary
of the systematic uncertainties in the analysis and their assigned values for each data set, as well
as additional information about the correlations between the uncertainties.

As the modified ABCD method for estimating the background
requires that the discriminating observables \ptmiss and \tpho are
independent, we propagate a systematic uncertainty
for any potential interdependence of these observables.
We select events in the \gjets and multijet CR
and separate events into the same A, B, C, and D bins defined for the signal
region. We compare the background yield in bin C predicted by the ABCD method
with the observed yield, and propagate the difference as a systematic uncertainty.
This systematic uncertainty is referred to as ``the closure'' in
Table~\ref{tab:syst_summary}. For the cases with neutralino proper decay length
smaller than $0.1$\m, this systematic uncertainty is relatively small, at
$4\%$ or less. For the cases with neutralino proper decay length larger than $0.1$\m,
the data yields in bin C of the CRs are small and are limited by statistical
uncertainty. As a result, a relatively large systematic uncertainty
of $90\%$ of the predicted background yield is propagated.

\begin{table*}[htb!]
  \topcaption{Summary of systematic uncertainties in the analysis. Also included are notes
  on whether each source affects signal yields (Sig) or background (Bkg) estimates, to which bins each uncertainty applies,
  and how the correlations of the uncertainties between the different data sets are treated. We assign different
  values for the uncertainty in the closure of the background prediction for short and long lifetime signal models.
  The column labeled 2017 includes both the \onepho{2017} and \twopho{2017} categories.
  \label{tab:syst_summary}}
  \centering

    \begin{scotch}{L{0.35\textwidth}C{0.075\textwidth}C{0.075\textwidth}C{0.075\textwidth}C{0.075\textwidth}C{0.18\textwidth}}
      Systematic uncertainty& Sig/Bkg & Bins & 2016 & 2017 & Correlation\\
      \hline
      Integrated luminosity & Sig & A,B,C,D & 2.5\% & 2.3\% & Uncorrelated \\
      Photon energy scale & Sig & A,B,C,D & 1\% & 2\% & Correlated\\
      Photon energy resolution & Sig & A,B,C,D & 1\% & 1\% & Correlated\\
      Jet energy scale & Sig & A,B,C,D & 1.5\% & 2\% & Correlated\\
      Jet energy resolution & Sig & A,B,C,D & 1.5\% & 1.5\% & Uncorrelated\\
      Photon time bias & Sig & A,B,C,D & 1.5\% & 1\% & Correlated\\
      Photon time resolution & Sig & A,B,C,D & 0.5\% & 0.5\% & Correlated\\
      Trigger efficiency & Sig & A,B,C,D & 2\% & $<$1\% & Uncorrelated\\
      Photon identification & Sig & A,B,C,D & 2\% & 3\% & Correlated \\
      Closure in bin C ($\ctau\leq0.1\m$) & Bkg & C & 2\% & 3.5\% & Correlated\\
      Closure in bin C ($\ctau>0.1\m$) & Bkg & C & 90\% & 90\% & Correlated\\
    \end{scotch}
\end{table*}

\section{Results and interpretation}
\label{sec:results}

Tables~\ref{tab:yields_twopho2016}~and~\ref{tab:yields_2017}
list the yields and postfit background predictions for the background-only fit
in each of the four bins of the 2016, \onepho{2017}, and \twopho{2017}
categories, respectively, for all the \tpho-\ptmiss bin boundaries used.
No statistically significant deviation from the background expectation is observed.
The search result is interpreted in terms of limits on the neutralino production
cross section for scenarios in the GMSB SPS8 signal model set.

\begin{table*}[htb!]
\centering
\topcaption{Observed number of events (\Nobs) and predicted background yields from the background-only fit (\Npostfit) in bins A, B, C, and D in data for the 2016 category and for the different \tpho and \ptmiss bin boundaries summarized in Table~\ref{tab:optimizedBinning}. In addition, the predicted postfit yields from the background-only fit not including bin C (\Npostfitmask) are provided as a test of the closure. Uncertainties in the \Npostfit and \Npostfitmask values are the postfit uncertainties. The propagation of the systematic uncertainties is handled during the fit and therefore they are included in the postfit uncertainties.}
\label{tab:yields_twopho2016}
\begin{scotch}{c c c c c c}
\multicolumn{6}{c}{2016 category} \\
\multicolumn{2}{L{0.22\textwidth}}{Bin boundary \newline [\tpho (ns), \ptmiss (\GeVns)]  } & A & B & C & D \\
\hline
\multirow{3}{*}{(0, 250)} & \Nobs & 16\,139 & 41 & 62 & 18\,826 \\
& \Npostfit & $16\,130 \pm 110$ & $47.5 \pm 4.8$ & $55.6 \pm 5.6$ & $18\,830 \pm 130$ \\
& \Npostfitmask & $16\,140 \pm 110$ & $41.0 \pm 6.5$ & $47.8 \pm 7.7$ & $18\,830 \pm 130$ \\[\cmsTabSkip]

\multirow{3}{*}{(1.5, 100)} & \Nobs & 33\,760 & 1302 & 1 & 5 \\
& \Npostfit & $33\,760 \pm 160$ & $1303 \pm 37$ & $0.29 \pm 0.28$ & $5.7 \pm 2.2$ \\
& \Npostfitmask & $33\,760 \pm 160$ & $1302 \pm 37$ & $0.19 \pm 0.21$ & $5.0 \pm 2.1$ \\[\cmsTabSkip]

\multirow{3}{*}{(1.5, 150)} & \Nobs & 34\,595 & 467 & 0 & 6 \\
& \Npostfit & $34\,600 \pm 170$ & $467 \pm 22$ & $0.08 \pm 0.08$ & $5.9 \pm 2.3$ \\
& \Npostfitmask & $34\,600 \pm 170$ & $467 \pm 22$ & $0.08 \pm 0.09$ & $6.0 \pm 2.3$ \\
\end{scotch}
\end{table*}

\begin{table*}[htb!]
\centering
\topcaption{Observed number of events (\Nobs) and predicted background yields from the background-only fit (\Npostfit) in bins A, B, C, and D in data for the \onepho{2017} (upper table) and \twopho{2017} (lower table) categories and for the different \tpho and \ptmiss bin boundaries summarized in Table~\ref{tab:optimizedBinning}. Additional details are described in the caption of Table~\ref{tab:yields_twopho2016}. }
\label{tab:yields_2017}
\begin{scotch}{c c c c c c}
\multicolumn{6}{c}{\onepho{2017} category} \\
\multicolumn{2}{L{0.22\textwidth}}{Bin boundary \newline [\tpho (ns), \ptmiss (\GeVns)]  } & A & B & C & D \\
\hline
\multirow{3}{*}{(0.5, 300)} & \Nobs & 458\,372 & 281 & 41 & 67\,655 \\
& \Npostfit & $458\,370 \pm 660$ & $281 \pm 15$ & $41.4 \pm 2.4$ & $67\,660 \pm 280$ \\
& \Npostfitmask & $460\,369 \pm 660$ & $281 \pm 16$ & $41.5 \pm 2.7$ & $67\,660 \pm 280$ \\[\cmsTabSkip]

\multirow{3}{*}{(1.5, 200)} & \Nobs & 524\,652 & 1364 & 1 & 332 \\
& \Npostfit & $524\,650 \pm 710$ & $1364 \pm 36$ & $0.9 \pm 0.8$ & $330 \pm 20$ \\
& \Npostfitmask & $524\,650 \pm 700$ & $1364 \pm 35$ & $0.9 \pm 1.0$ & $330 \pm 20$ \\[\cmsTabSkip]

\multirow{3}{*}{(1.5, 300)} & \Nobs & 525\,694 & 322 & 0 & 333 \\
& \Npostfit & $525\,690 \pm 700$ & $322 \pm 17$ & $0.19 \pm 0.21$ & $330 \pm 20$ \\
& \Npostfitmask & $525\,690 \pm 700$ & $322 \pm 17$ & $0.20 \pm 0.24$ & $330 \pm 20$ \\

\hline
\noalign{\vskip6pt}

\multicolumn{6}{c}{\twopho{2017} category} \\[\cmsTabSkip]
\multirow{3}{*}{(0.5, 150)} & \Nobs & 21\,640 & 362 & 56 & 3201 \\
& \Npostfit & $21\,640 \pm 140$ & $364 \pm 17$ & $54.0 \pm 3.0$ & $3200 \pm 60$ \\
& \Npostfitmask & $21\,640 \pm 140$ & $362 \pm 18$ & $53.6 \pm 3.3$ & $3200 \pm 60$ \\[\cmsTabSkip]
\multirow{3}{*}{(0.5, 200)} & \Nobs & 21\,863 & 139 & 24 & 3233 \\
& \Npostfit & $21\,860 \pm 140$ & $142 \pm 11$ & $21.1 \pm 1.7$ & $3240 \pm 60$ \\
& \Npostfitmask & $21\,860 \pm 140$ & $139 \pm 11$ & $20.6 \pm 1.8$ & $3230 \pm 60$ \\[\cmsTabSkip]
\multirow{3}{*}{(1.5, 150)} & \Nobs & 24\,824 & 418 & 0 & 17 \\
& \Npostfit & $24\,820 \pm 150$ & $420 \pm 20$ & $0.25 \pm 0.28$ & $16.7 \pm 4.4$ \\
& \Npostfitmask & $24\,820 \pm 150$ & $420 \pm 20$ & $0.29 \pm 0.36$ & $17.0 \pm 4.4$ \\[\cmsTabSkip]
\multirow{3}{*}{(1.5, 200)} & \Nobs & 25\,079 & 163 & 0 & 17 \\
& \Npostfit & $25\,080 \pm 150$ & $163 \pm 12$ & $0.11 \pm 0.12$ & $16.9 \pm 4.4$ \\
& \Npostfitmask & $25\,080 \pm 150$ & $163 \pm 12$ & $0.11 \pm 0.14$ & $17.0 \pm 4.4$ \\

\end{scotch}
\end{table*}

\begin{figure}[hbtp]
\centering
\includegraphics[width=1.0\columnwidth]{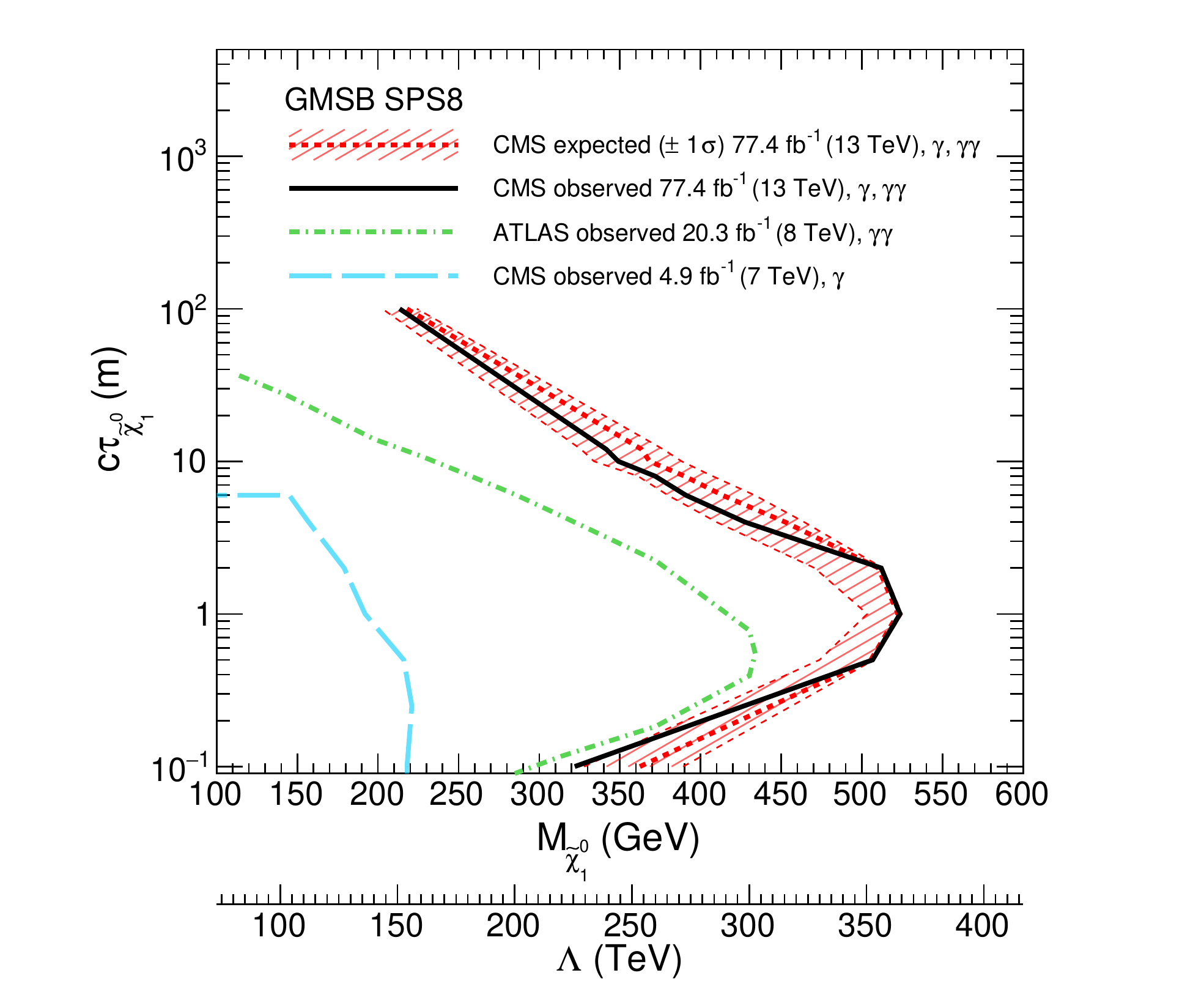}
\caption{The 95\% \CL exclusion contours for the GMSB neutralino production
cross section, shown as functions of the neutralino mass, or equivalently the
SUSY breaking scale, $\Lambda$, in the GMSB SPS8 model, and the neutralino proper
decay length, $\ctau_{\PSGczDo}$.}
\label{fig:Limits2D}
\end{figure}

The modified frequentist criterion \CLs~\cite{Junk:1999kv,Read:2002hq,ATLAS:2011tau}
with the profile likelihood ratio test statistic determined by toy experiments
is used to evaluate the observed and expected limits at 95\%~confidence level (\CL)
on the signal production cross sections.
The limits are shown in Fig.~\ref{fig:Limits2D} as functions of the mass of the neutralino NLSP $\PSGczDo$
(linearly related to the SUSY breaking scale, $\Lambda$) and the proper decay length of the
neutralino.
The two-photon category (2016 and \twopho{2017}) and the one-photon category (\onepho{2017}) are
complementary as the sensitivity at small proper decay length is better
for the 2016 and \twopho{2017} categories because of the extra background suppression from requiring
two photons, while the sensitivity at large proper decay lengths is better for the \onepho{2017} analysis
because of the significantly improved signal acceptance from the dedicated displaced single-photon
trigger. As a result, the sensitivity to signal models with proper lifetimes greater than the
ECAL timing resolution for a single photon candidate is improved compared to previous results.
For the neutralino proper decay lengths \ctau of $0.1$, $1$, $10$, and $100$\m,
masses up to about 320, 525, 360, and 215\GeV are excluded at 95\% \CL, respectively.

\section{Summary}
\label{sec:conclusion}

A search for long-lived particles that decay to a photon and a weakly interacting particle
has been presented. The search is based on proton-proton collisions at a center-of-mass
energy of 13\TeV collected by the CMS experiment in 2016--2017.
The photon from this particle's decay would enter the electromagnetic calorimeter at
non-normal impact angles and with delayed times, and
this striking combination of features is exploited to suppress backgrounds.
The search is performed using a combination of the 2016 and 2017 data sets, corresponding to a total
integrated luminosity of 77.4\fbinv. Both single-photon and diphoton event samples are used
for the search, with each sample providing a complementary sensitivity
at larger and smaller long-lived particle proper decay lengths, respectively.
The results are interpreted in the
context of supersymmetry with gauge-mediated supersymmetry breaking, using the SPS8
benchmark model. For neutralino proper decay lengths of 0.1, 1, 10, and 100\m,
masses up to about 320, 525, 360, and 215\GeV are excluded at 95\% confidence level, respectively.
The previous best limits are extended by one order of magnitude in the neutralino proper decay
length and by 100\GeV in the mass reach.

\begin{acknowledgments}
We congratulate our colleagues in the CERN accelerator departments for the excellent performance of the LHC and thank the technical and administrative staffs at CERN and at other CMS institutes for their contributions to the success of the CMS effort. In addition, we gratefully acknowledge the computing centers and personnel of the Worldwide LHC Computing Grid for delivering so effectively the computing infrastructure essential to our analyses. Finally, we acknowledge the enduring support for the construction and operation of the LHC and the CMS detector provided by the following funding agencies: BMBWF and FWF (Austria); FNRS and FWO (Belgium); CNPq, CAPES, FAPERJ, FAPERGS, and FAPESP (Brazil); MES (Bulgaria); CERN; CAS, MoST, and NSFC (China); COLCIENCIAS (Colombia); MSES and CSF (Croatia); RPF (Cyprus); SENESCYT (Ecuador); MoER, ERC IUT, PUT and ERDF (Estonia); Academy of Finland, MEC, and HIP (Finland); CEA and CNRS/IN2P3 (France); BMBF, DFG, and HGF (Germany); GSRT (Greece); NKFIA (Hungary); DAE and DST (India); IPM (Iran); SFI (Ireland); INFN (Italy); MSIP and NRF (Republic of Korea); MES (Latvia); LAS (Lithuania); MOE and UM (Malaysia); BUAP, CINVESTAV, CONACYT, LNS, SEP, and UASLP-FAI (Mexico); MOS (Montenegro); MBIE (New Zealand); PAEC (Pakistan); MSHE and NSC (Poland); FCT (Portugal); JINR (Dubna); MON, RosAtom, RAS, RFBR, and NRC KI (Russia); MESTD (Serbia); SEIDI, CPAN, PCTI, and FEDER (Spain); MOSTR (Sri Lanka); Swiss Funding Agencies (Switzerland); MST (Taipei); ThEPCenter, IPST, STAR, and NSTDA (Thailand); TUBITAK and TAEK (Turkey); NASU and SFFR (Ukraine); STFC (United Kingdom); DOE and NSF (USA).

\hyphenation{Rachada-pisek} Individuals have received support from the Marie-Curie program and the European Research Council and Horizon 2020 Grant, contract Nos.\ 675440, 752730, and 765710 (European Union); the Leventis Foundation; the A.P.\ Sloan Foundation; the Alexander von Humboldt Foundation; the Belgian Federal Science Policy Office; the Fonds pour la Formation \`a la Recherche dans l'Industrie et dans l'Agriculture (FRIA-Belgium); the Agentschap voor Innovatie door Wetenschap en Technologie (IWT-Belgium); the F.R.S.-FNRS and FWO (Belgium) under the ``Excellence of Science -- EOS" -- be.h project n.\ 30820817; the Beijing Municipal Science \& Technology Commission, No. Z181100004218003; the Ministry of Education, Youth and Sports (MEYS) of the Czech Republic; the Lend\"ulet (``Momentum") Program and the J\'anos Bolyai Research Scholarship of the Hungarian Academy of Sciences, the New National Excellence Program \'UNKP, the NKFIA research grants 123842, 123959, 124845, 124850, 125105, 128713, 128786, and 129058 (Hungary); the Council of Science and Industrial Research, India; the HOMING PLUS program of the Foundation for Polish Science, cofinanced from European Union, Regional Development Fund, the Mobility Plus program of the Ministry of Science and Higher Education, the National Science Center (Poland), contracts Harmonia 2014/14/M/ST2/00428, Opus 2014/13/B/ST2/02543, 2014/15/B/ST2/03998, and 2015/19/B/ST2/02861, Sonata-bis 2012/07/E/ST2/01406; the National Priorities Research Program by Qatar National Research Fund; the Ministry of Science and Education, grant no. 3.2989.2017 (Russia); the Programa Estatal de Fomento de la Investigaci{\'o}n Cient{\'i}fica y T{\'e}cnica de Excelencia Mar\'{\i}a de Maeztu, grant MDM-2015-0509 and the Programa Severo Ochoa del Principado de Asturias; the Thalis and Aristeia programs cofinanced by EU-ESF and the Greek NSRF; the Rachadapisek Sompot Fund for Postdoctoral Fellowship, Chulalongkorn University and the Chulalongkorn Academic into Its 2nd Century Project Advancement Project (Thailand); the Nvidia Corporation; the Welch Foundation, contract C-1845; and the Weston Havens Foundation (USA).
\end{acknowledgments}

\bibliography{auto_generated}
\cleardoublepage \appendix\section{The CMS Collaboration \label{app:collab}}\begin{sloppypar}\hyphenpenalty=5000\widowpenalty=500\clubpenalty=5000\vskip\cmsinstskip
\textbf{Yerevan Physics Institute, Yerevan, Armenia}\\*[0pt]
A.M.~Sirunyan$^{\textrm{\dag}}$, A.~Tumasyan
\vskip\cmsinstskip
\textbf{Institut f\"{u}r Hochenergiephysik, Wien, Austria}\\*[0pt]
W.~Adam, F.~Ambrogi, T.~Bergauer, J.~Brandstetter, M.~Dragicevic, J.~Er\"{o}, A.~Escalante~Del~Valle, M.~Flechl, R.~Fr\"{u}hwirth\cmsAuthorMark{1}, M.~Jeitler\cmsAuthorMark{1}, N.~Krammer, I.~Kr\"{a}tschmer, D.~Liko, T.~Madlener, I.~Mikulec, N.~Rad, J.~Schieck\cmsAuthorMark{1}, R.~Sch\"{o}fbeck, M.~Spanring, D.~Spitzbart, W.~Waltenberger, C.-E.~Wulz\cmsAuthorMark{1}, M.~Zarucki
% ...
\vskip\cmsinstskip
\dag: Deceased\\
1:  Also at Vienna University of Technology, Vienna, Austria\\
% ...
\end{sloppypar}
\end{document}